\documentclass[modulo,mathlines,switch,running,columnwise,pagewise,desactivate]{aa}
\usepackage{graphicx}
\usepackage{natbib}
\usepackage{amsmath}
\usepackage{xcolor}

\title{Secondary impact debris in the Didymos system: what could be observed by Hera?}
\author{K.Langner
\inst{1,}\inst{2}
\and
E. Martellato \inst{3}
\and
R. Luther \inst{4}
\and
F. Marzari \inst{5}
\and
A. Rossi \inst{1}
}

\institute{
   IFAC-CNR, Via Madonna del Piano 10, 50019 Sesto Fiorentino, Italy  \and 
      Astronomical Observatory Institute, Faculty of Physics and Astronomy, Adam Mickiewicz University, Poznań, Poland
      \and
      Osservatorio Astronomico di Padova, Padova, Italy
      \and
    Museum für Naturkunde – Leibniz Institute for Evolution and Biodiversity Science, Berlin, Germany\and
    Dipartimento di Fisica, Universit\`a di Padova, Italy
   }

\date{Received 17.02.2025}


\begin{document}

\abstract{In 2022 the DART spacecraft impacted the asteroid Dimorphos, the small moonlet of the asteroid Didymos, ejecting dust, rocks and boulder into the space around both asteroids. A part of those ejecta can re-impact the surface of the two asteroids. At the end of 2026, ESA's Hera mission \citep{Michel} will arrive to the system to analyse in detail the binary system and the impact consequences.}
{We investigate the effects of low—velocity impacts of rocks and boulders, originally released after the DART impact, on the surface of Didymos and the dynamics of dust particles released by those impacts. We determine if any of those effects can be observed by the Hera mission.}{The iSALE-2D shock physics code was used to simulate the re-impacts of boulders on the surface of the asteroid. To model the dynamics of the boulders, we used a numerical model that includes the gravity of non-spherical Didymos and Dimorphos, the solar gravity, and the radiation pressure.}{The sesquinary impacts can result in small, shallow craters on the surface of Didymos. For the given low impact speeds, the ejected mass depends mostly on the boulder mass. Ejection speeds range from $\sim 10$ \% to $\sim 80$ \% of the impact speed. The majority of the ejected dust falls back covering a large area of the surface, mostly at low/medium latitudes. Less than 20 \% of the ejected dust is escaping the system after a few days. The space surrounding the asteroids becomes free from dust after 15-30 days following each sesquinary impact.}{}
\maketitle
\section{Introduction}

The impact of the DART probe on Dimorphos \citep{CHENG2015}, the small moon 
of the binary 65803 Didymos asteroid system \citep{NatureDaly, NatureCheng, NatureDotto}, caused a large  number of ejecta fragments of variable dimensions, from micron--sized dust particles to meter--sized boulders, to be launched from the impact site with different ejection velocities. Their subsequent evolution is characterized by a variety of dynamical paths \citep{langner2024,rossi2022,Richardson_2024,Ferrari}, determined by the gravity field of the binary asteroidal system, the Sun tidal force, and the solar radiation pressure. The dominant evolutionary channels are re-impact on Dimorphos or impact on Didymos or hyperbolic escape from the system into heliocentric orbits. 

In some Hubble Space Telescope images, \cite{Jewitt} identified a population of low velocity boulders ejected by the DART impact, prone to remain trapped within the binary system. \cite{langner2024} performed an analysis of the evolution of a sample population of boulders similar to those observed by \citep{Jewitt}. They numerically simulated the evolution of large boulders ejected by the DART impact in the complex environment of the binary asteroidal system. Their analysis pointed out that the vast majority of the boulders tend to re-impact against Didymos due to its larger mass and cross section. These events are defined in literature as sesquinary impacts (see, e.g., \cite{sesqui}). In more details, Langner et al. found that after an initial peak in the frequency of re-impacts on the two asteroids, which lasts approximately 30--40 days, the re-impacts then continue at a steady low rate for an extended period of time, with some boulders possibly re-impacting after $3.5 \div 4$ years.  The initial high frequency of sesquinary impacts lifts from the surface of the asteroid additional dust particles, which may refill the dust tail developed immediately after the DART impact. This second generation of dust may explain the extended period of time during which a tail behind the Didymos system has been observed \citep{Moreno2023,Ferrari}. 
In addition, these sesquinary impacts are expected to alter the surface features (e.g., by increasing the local density of boulders and shallow craters) and the local composition by exposing lower strata of the soil of the two asteroids. To correctly interpret the images that will be taken by the Hera mission \citep{Michel}, we have to take into account also the effects of such sesquinary impacts, by modelling them in details.

In this paper, our goal is to study the evolution of the ejecta produced by sesquinary collisions, on the surface of Didymos. Note that we concentrate on the primary asteroid since, as mentioned, most of the re-impacts happen on this body. Moreover, due to their higher velocities, they are prone to generate consequences that are more likely observed by Hera. In particular, we investigate for how long the ejecta hover on the surface of the body, the fraction of ejecta which re--impacts on either one of the asteroids, and the percentage that directly escapes from Didymos or orbits for a while in the binary system, before definitively escaping, posing a possible threat to the Hera mission.

Our modelling is based on two sequential steps. In the first one, we model by means of the shock physics code iSALE \citep{collins2016} the typical sesquinary impacts on the surface of Didymos. The impact parameters are derived from our previous long--term simulations of the DART ejecta evolution \citep{langner2024}. The iSALE outputs provides the initial conditions for the ejected fragments (mass and ejection velocities and angles), which are then used as input for the LICEI code \citep{rossi2022}, representing the second step. It integrates in time the dynamical evolution of these second generation of fragments. In this way we can predict their flight times before re--impact and their final distribution on the surface of the asteroids.

\section{Numerical models description}

\subsection{Sesquinary impact simulations with iSALE}

The iSALE-2D shock physics code, Dellen version, (e.g., \citep{collins2016}) was used to model sesquinary impacts on the Didymos surface, in order to compute the ejecta properties. iSALE is a multi-material and multi-rheology code developed from the SALE hydrocode \citep{amsden1980}. It includes a number of models, such as (i) elasto-plastic constitutive, (ii) porous-compaction, and (iii) dilatancy models, various equations of state (EoS), and multiple materials, allowing to simulate the physics of the shock waves propagation and the cavity opening on a specific planetary body (e.g., \citep{collins2004, collins2011, kai2006}). A benchmarking campaign to validate iSALE against other hydrocodes has been carried out and summarized in \citep{pierazzo2008, robert2022, STICKLE2020}.\\

The impact was modelled in a cylindrically symmetric grid. The target -- Didymos -- has a size much larger than the projectile sizes (mean radius of about 350 m against a radius of 0.5 m for the impactors used in this work). For this reason, we adopted a planar semi-infinite space as target instead of a sphere, reducing significantly the computation time. Any test adopting instead spherical colliding bodies was too highly time consuming and the crater formation did not evolve enough to appreciate differences with respect to the semi-infinite half space.\\
The grid is made up by a high resolution zone of 100 × 80 cells, surrounded by a low resolution zone defined by the rule that one cell is 1.05 times the previous one, in order to prevent reflections at the boundaries. The target area is in total 3.4 m × 4.1 m and 34 m × 41 m, respectively in two tested projectile sizes (10 cm and 1 m as described below). This is an Eulerian grid, where material advects through adjacent cells. Furthermore, Lagrangian tracers were implemented in the models, and were initially distributed in the centre of each cell. These are massless points that stand for cylindrical symmetric masses, and therefore the higher the distance from the symmetric axis the larger are the masses they represent \citep{collins2016}. Once the excavation flow begins, the Lagrangian tracers move with the velocity field during crater evolution, recording material displacement and allowing to derive ejecta motion. Upon ejection, tracers were analysed to derive the mass that they represent, the ejection speed, and the ejection angle with respect to the target surface. Following the approach shown by \citep{robert2018}, an initial ejection altitude needs to be defined above the surface in order to both avoid the large variations established by the pressure gradient close to the surface and to catch also the slow ejecta. The adopted ejection criterion was defined as an ejection altitude that needs to be exceeded by tracers, here 2 cm and 20 cm for the projectile sizes of 10 cm and 1 m, respectively (cf. Tab.~\ref{tab::isale-input}).\\

The inner structure and composition of the two bodies of the asteroidal system are still poorly constrained. Both Didymos and Dimorphos are expected to be similar in composition, since the two bodies possibly share the same origin \citep{barnouin, raducan23_inpress, walsh-jac}).
Spectral studies suggested an S-type silicaceous composition, with a possible L-type ordinary chondritic composition revealed before the DART impact (e.g., \citep{bagnulo2023, polishook2023}). In a number of modelling studies reproducing the DART impact, a basaltic composition was used for Dimorphos (e.g., \citep{raducan2020, raducan2022, robert2022}). Basaltic asteroids are shaped through re-crystallization of melted rocks, which in turns were likely formed after the breakup of large differentiated bodies (e.g., \citep{nesvorny2008}). Indeed, most of the basaltic asteroids smaller than 1-2 km in size in the inner Main Belt are hypothesized to be (4) Vesta family members, while the outer Main belt basaltic objects could be attributed to other differentiated bodies like (1459) Magnya (e.g., \citep{Moreno2023, hardersen}).
The impactors are expected to be the boulders produced by the DART impact on Dimorphos. Therefore, a basaltic composition could be a genuine choice for both the projectile and the target of this study.\\

The projectile was simplified as an homogeneous sphere, set to a resolution of 10 cells per projectile radius (cppr), which implies a cell size of 5 × 10\textsuperscript{-3} m and 5 × 10\textsuperscript{-2} m, respectively for the smaller and larger dimensions. The projectile diameter was set to 10 cm and 1 m (the boulder diameter), and its velocity to three proxy values: 15, 45, and 65 cm/s. These velocities were derived from the dynamical simulations of re-impacting boulders on the Didymos--Dimorphos pair and are representative of the range of velocities obtained in \citep{langner2024} (see Fig. 4 in that reference). Although iSALE is a shock-physics code aiming at simulating hypervelocity impacts (i.e., events with speed of the order of km/s, e.g., \citep{melosh}), it has recently proven to correctly simulate impacts also at lower impact speeds down to few m/s \citep{laetitia2023}.\\
Basalt was described by means of the Tillotson EoS, with reference density of 2650 kg/m\textsuperscript{3} \citep{tillo1962}, whereas the shear strength was modelled using simple models describing either granular-like material \citep{drucker1952}, or rock-like material \citep{LUNDBORG1967}, referred here-after as DRPR and LUND, respectively. The cohesive strength of the damaged target material (at zero pressure) was set to 10 Pa, proposed as inner cohesion for Didymos on the basis of observations and comparison with other asteroids \citep{barnouin,raducan23_inpress} and adopted in other modelling studies of the DART impact (e.g., \citep{robert2022, ferrari2022}). the small size of Didymos, together with the assumed cohesion of 10 Pa \citep{raducan23_inpress}, implies that the impacts occur in a strength regime. From previous numerical studies, it was advanced a value of 1 to 4 Pa as cohesion threshold for the strength-gravity transition \citep{raducan23_inpress, stickle2022}. The internal friction coefficient was fixed to a value of 0.6, which is typical for basalt at ambient temperature \citep{schultz1993}.\\
For the projectile and target, a porosity was also considered and described by the $\varepsilon-\alpha$ porosity model \citep{kai2006, collins2011}. We tested two cases: (i) same porosity, set to $40\%$, (ii) different porosities, set to $10\%$ and $60\%$, respectively for the projectile and target. A bulk porosity range between 40 and $60\%$ was suggested by \citep{britt-consol2001} for rubble piles, and a $30-40\%$ of porosity represents a threshold for compaction cratering \citep{HOUSEN201872}. In this work, we therefore considered these two values to assess the impacts of the test particles. We used the values adopted by \citep{raducan2022} as input parameters for the $\varepsilon-\alpha$ porosity model.\\
The input model parameters were summarized in Tab.~\ref{tab::isale-input}.\\

\subsection{Dynamics of the ejecta modelled with LICEI}

The sesquinary impacts raised new dust particles, which evolve in time, and may (i) survive for a while in orbit around the body creating a sort of haze, or (ii) be ejected out of the system contributing to the prolongation of the post-impact tail. To model the dynamics of such dust particles ejected after sesquinary impacts on the surface of Didymos (or Dimorphos), we used the LICEI code \citep{rossi2022,langner2024}.
The code exploits the Radau numerical integrator \citep{radau1985} to follow the trajectories in a quickly varying gravitational field similar to the one of the binary asteroid system. To speed up computations, the approximate MacCullagh formula \citep{murray2000} was adopted to describe the gravity field of each individual non-spherical asteroid rather than using the computationally expensive polyhedral model. This formula uses the moments of inertia of a body to compute the gravitational potential
\begin{equation}
V= - \frac{GM}{r} -
 \frac{G(A+B+C-3I)}{2 r^3},
\end{equation}
where $A,B,C$ are principal axes moments of inertia of the body, $r$ is a distance to the center of mass of the body and
\begin{equation}
I = \frac{(Ax^2+By^2+Cz^2)} {r^2}.
\end{equation}
We took the triaxial ellipsoid approximation for the shape with semi-axes $a,b,c$, so the inertia moments are given by the equations
\begin{eqnarray*}
A & = & \frac{4}{15} \pi \rho abc (b^2 + c^2) \\
B & = & \frac{4}{15}\pi \rho abc (a^2 + c^2) \\
C & = & \frac{4}{15}\pi \rho abc (a^2 + b^2).
\label{Maccullagh3}
\end{eqnarray*}

The values of $a,b,c$ for both asteroids were derived from the shape model produced by means of the observations of the DART mission \citep{NatureDaly}, and reported in the SPICE kernels. The numerical integration of the orbits of the fragments was performed in a non-rotating reference frame centred on the Didymos--Dimorphos barycentre.

The solar tidal force, which is intended as the difference between the Sun gravitational force acting on the fragment and the one on Didymos, was added using the position and velocity vectors given by the SPICE kernels.

The solar radiation pressure was added to the total force acting on each particle using the usual formulation 

\begin{equation}
\boldsymbol {F}_{RP}=  \frac{SA} {c} Q_{PR} \boldsymbol {s},
\end{equation}

where $S$ is the solar radiation flux density at the heliocentric distance of the body, $A$ is the geometrical cross section of the fragment, and $Q_{PR}$ is a dimensionless coefficient determining the amount of radiation that is either reflected or absorbed and then re-emitted \citep{burns1979}. The versor $\boldsymbol {s}$ was directed in the anti-solar direction, which was taken out from the SPICE kernels.
\\

\section{Results: impact simulations}

\begin{table*}[h!]
\caption{Results of crater sizes of iSALE models for the two used strength models DRPR and LUND. In the first lines of each section (marked with ${\dagger}$ sign) both the projectile's and target's porosity are set to 40 \%, whereas in all the other cases the projectile's and target's porosity are 10\% and 60\%, respectively. The table columns report: (i) projectile diameter L, (ii) impact velocity $v_i$, (iii) crater diameter at the pre-impact surface $D_{surf}$, (iv) and (v) crater diameter $D_c$ and depth $d_c$, respectively, (vi) depth-to-diameter ratio $d_c/D_c$.\\}
\label{tab::iSALE}
\centering
\begin{tabular}{c c c c c c}
\hline\hline
L $[m]$  & $v_i [cm/s]$ & $D_{surf} [m]$ & $D_c [m]$ & $d_c [m]$ & $d_c/D_c$\\
\hline
DRPR\\
0.1 ${\dagger}$ & 45 & 0.170  & 0.200 & 0.030 & 0.152\\

0.1 & 45 & 0.256  & 0.260 & 0.056 & 0.217\\
0.1 & 65 & 0.283  & 0.340 & 0.080 & 0.235\\

1 & 45 & 2.192  & 2.500 & 0.538 & 0.215\\
1 & 65 & 2.532  & 2.800 & 0.659 & 0.235\\
\\
\hline
LUND\\
0.1 ${\dagger}$ & 45 & 0.187  & 0.200 & 0.032 & 0.162\\

0.1 & 45 & 0.250  & 0.260 & 0.048 & 0.184\\
0.1 & 65 & 0.290  & 0.290 & 0.090 & 0.310\\

1 & 45 & 2.279  & 2.400 & 0.482 & 0.201\\
1 & 65 & 2.589  & 2.900 & 0.788 & 0.272\\
\\
\hline

\end{tabular}
\vspace{1ex}

{\raggedright $\dagger$ Both projectile's and target's porosity is 40\%. The same setup for the 1 m-size impact, is not reported, due to the high computational cost.\par}

\end{table*}

\begin{figure}[h!]
\centering
\includegraphics[width=\hsize]{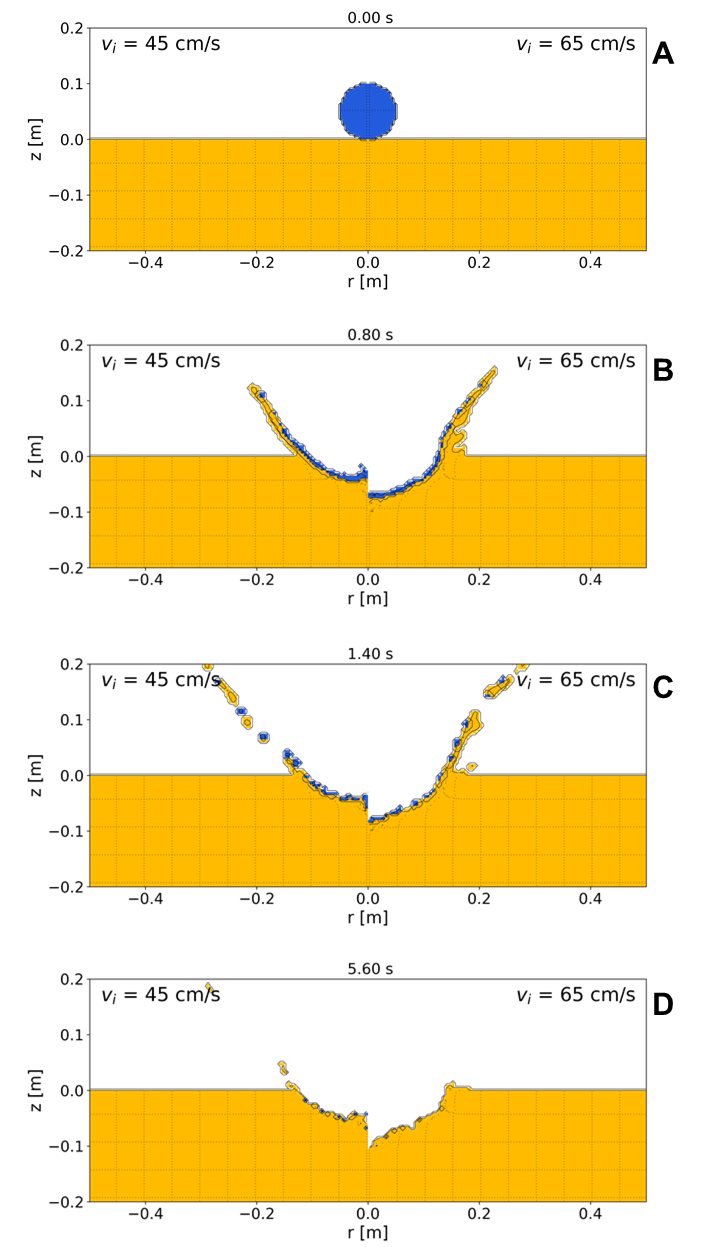}
\caption{Snapshots of crater formation at 0 (A), 0.8 (B), 1.4 (C), and 5.6 (D) seconds after the impact. Each plot is divided in two panels: the left panels show the results for an impact speed of $45$ cm/s, while the right panels refer to an impact speed of $65$ cm/s. The projectile is 10 cm in diameter. The colours refer to the different materials (blue for the projectile, and yellow for the target), which differ for the porosity.}
\label{fig:mod_crater_formation}
\end{figure}

The main goal of the simulations with the iSALE code is to derive the initial mass and velocity of the ejecta fragments of typical sesquinary impacts after the DART impact, which will then serve as initial conditions for the dynamical model propagating their trajectories in time. Tab.~\ref{tab::iSALE} details the crater morphology outputs for the different projectile sizes and velocities considered. As mentioned above, two material models are adopted, and the results are separately given in Tab.~\ref{tab::iSALE}.
In Fig.~\ref{fig:mod_crater_formation} we show the evolution of a crater produced by an impact of a $10\%$ basaltic projectile, having a diameter of 10 cm, on a regolith-like target. The impact speeds are $45$ cm/s and $65$ cm/s on the left and right panels of each plot, respectively. The series of plots from top to bottom shows the crater evolution at different times, and precisely 0.8, 1.4, and 5.6 sec, respectively, in addition to the contact point (t=0 s). In Fig.~\ref{fig:mod_crater_formation}B the cavity is expanding, with the projectile not completely vaporized. It is lining both the wall and floor of the expanding cavity, and it is well mixed to the ejecta material. In the snapshot Fig.~\ref{fig:mod_crater_formation}C the transient cavity appears, and in snapshot D we provide the final stages of the crater formation ($T_{fin}$=5.60 s). While for the simulation shown on the right ($v$ = 65 cm/s) the crater is completely formed, for the lower speed case on the left ($v$ = 45 cm/s) the ejecta material is still moving around the crater, likely due to the very low surface gravity. Nevertheless, also in this last case, the crater size can be considered in its final shape, having formed in the strength regime. We also verified that, even though some material can still be floating around the impact site, the crater diameter and depth have not varied in the last $0.3$ seconds of the evolution (i.e., from ($T_{fin} - 0.3$) sec up to $T_{fin}$).
\\

\begin{figure}[h!]
\centering
\includegraphics[width=\hsize]{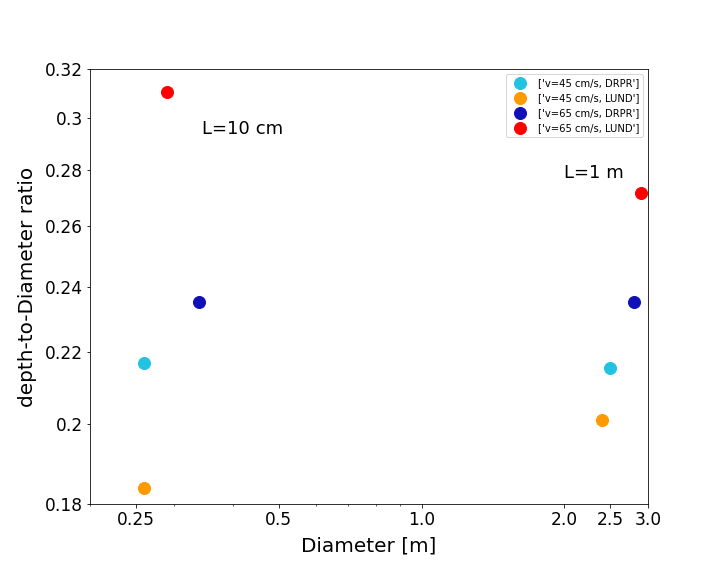}
\caption{Plot of the depth-to-diameter ratio of the simulated craters (see also the 6th column in Tab.~\ref{tab::iSALE}). The points on the left side refer to an impact of a projectile with a diameter of 10 cm while those on the right side to a 1 m projectile, respectively. The different colours represent variations in impact velocity and material models, as shown in the the plot legend.}
\label{fig:dD}
\end{figure}

The last snapshot in Fig.~\ref{fig:mod_crater_formation} highlights the lower depth-to-diameter ratio $d_c$/$D_c$ obtained in the 45 $cm/s$ simulation case. This is also noticed in all the tested cases  run at the lower speed (cf. Fig.~\ref{fig:dD} and Tab.~\ref{tab::iSALE}). One exception to these findings is given by the simulation runs at 15 cm/s for both the 10 cm and 1 m-sized projectiles. We find that for this specific value of speed, no crater forms and projectile is destroyed on the surface.\\

\begin{figure*}[h!]
\centering
\includegraphics[width=1.\textwidth]{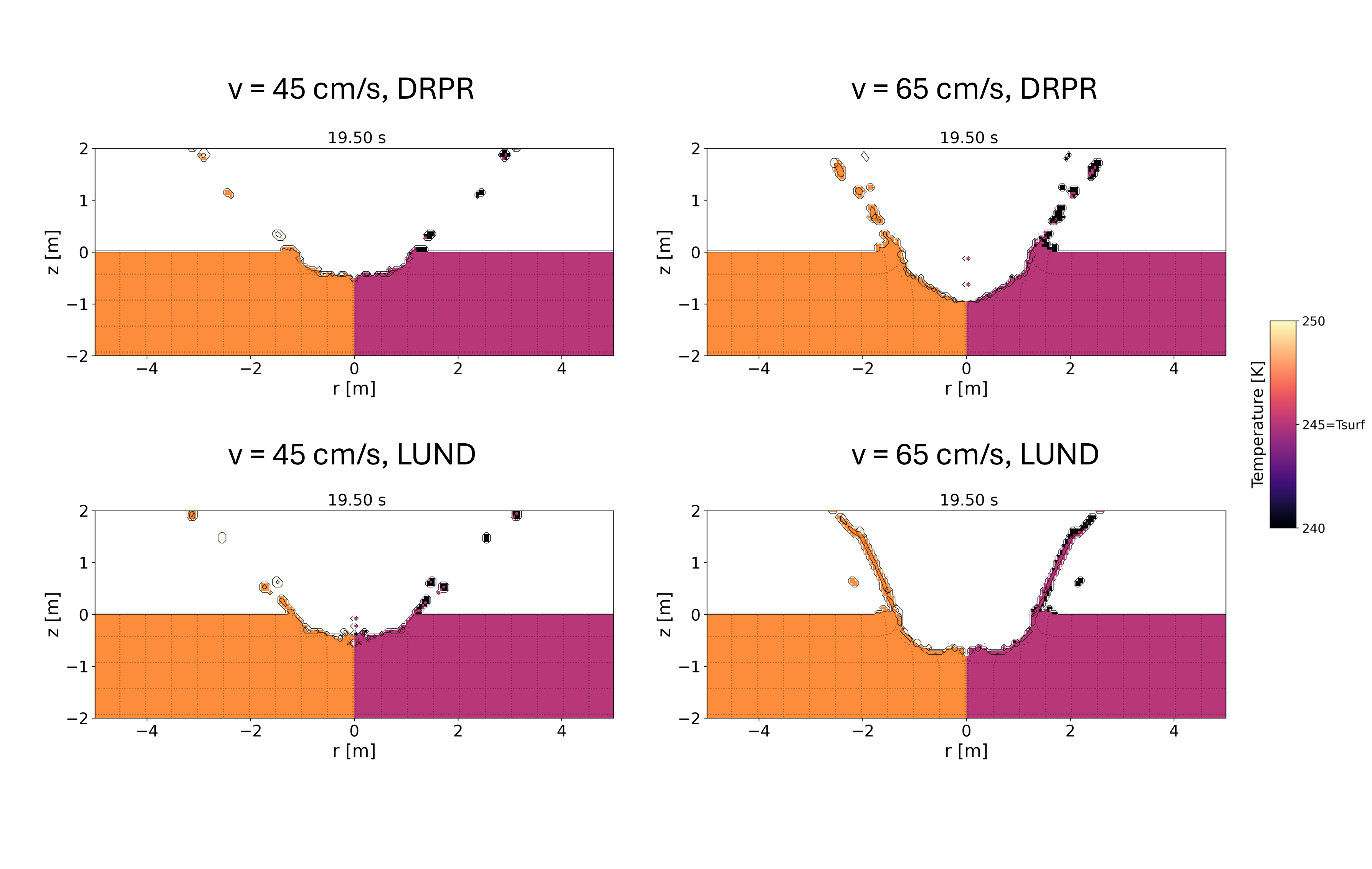}
\caption{Snapshots at 19.50 s of a 10\% basaltic projectile of 1 m in diameter, impacting on a 60\% target. The left panels show the material distribution, where white is the projectile and orange is the target. A different behaviour of the projectile and target materials can be noticed by the different ejection angle of the two curtains. The right panels of each plot show the temperature distribution, according to the colour legend. A variation in the temperature is observed only in the ejecta curtain.}
\label{fig:mod_feb_1m}
\end{figure*}

As mentioned above, in this work, we consider two different strength models for the target, i.e. the DRPR and LUND models, in order to cover a variety of cases and appraise how the ejecta evolution and the final crater morphology can be influenced by the surface properties. The DRPR rheological model has been used in a number of modelling investigations to describe the behaviour of regolith-like material during crater formation (e.g., \citep{prieur2017}). Being the strength linearly dependent on the pressure in this model, we also evaluate crater evolution by adopting the LUND model, for which the maximum pressure is instead asymptotically reached (e.g., \citep{robert2022}).

On average, we obtain a $d_c$/$D_c$ of about 0.2, a value closer to the one of simple craters on the terrestrial planets, rather than on small bodies, for which the average $d_c$/$D_c$ is lower and of the order of 0.12--0.15 (e.g., \citep{NOGUCHI} and references within). However, the range of $d_c$/$D_c$ measured on the asteroids visited by a spacecraft can reach values as large as 0.35 on Vesta \citep{VINCENT2014}. For a given crater class (e.g., simple crater), the $d_c$/$D_c$ represents a measure of the projectile and target properties and subsequent evolution of the craters. The latter is not accounted on these numerical models, being representative of fresh structures. For the cases tested in this study, different simulation outputs are obtained when using either the DRPR or LUND strength models (cf. Tab.~\ref{tab::iSALE}). For both the projectile sizes, there is a general increase of $d_c$/$D_c$ with velocity, i.e. passing from 45 cm/s to 65 cm/s. However, as highlighted in Fig.~\ref{fig:dD}, such an increase in the $d_c$/$D_c$ is qualitatively different for the two rheological models, and in particular much larger for the LUND one. Indeed, we notice an increase of $\sim$8\% when adopting the DRPR, while of $>$25\% when instead adopting the LUND one. This variation can be mainly ascribed to the crater depth, which undergoes an increment of 20--30\% and 40--45\% with increasing velocity, respectively for the two rheological models (cf. Fig.~\ref{fig:dD}). The different morphology of the cavity obtained for varying impact speeds and rheological models can be appraised also in Fig.~\ref{fig:mod_feb_1m}, which displays snapshots of a simulation at $t\sim19.5$ s for two different impact velocities (45 cm/s and 65 cm/s) of a 1 m size boulder on a target described by either the DRPR or the LUND rheological models. This might possibly be ascribed to the fact that DRPR is relatively stronger with respect to the LUND rheology due to the shape of the yield envelope. The latter thus causes the craters to form deeper in the 65 cm/s case.

The variation of the $d_c$/$D_c$ with increasing projectile size (from 10 cm to 1 m) has a more complex behaviour (keeping fixed the impact speed). Indeed it remains almost constant when adopting the DRPR model, whereas, when the LUND model was instead used, it exhibits an increase for the 45 cm/s impact velocity case and a decrease for the 65 cm/s one.

\begin{figure}[h!]
\centering
\includegraphics[width=\hsize]{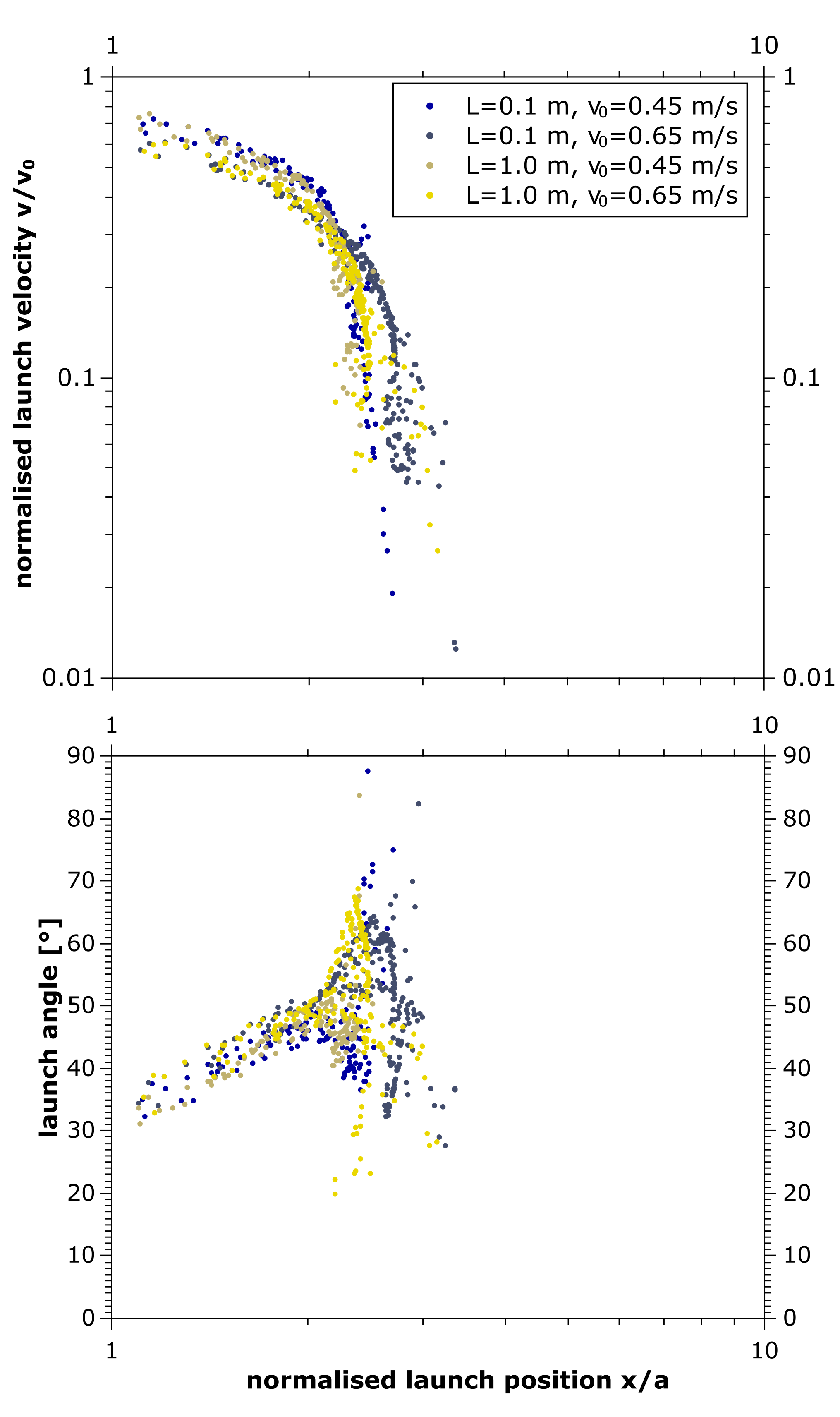}
\caption{Plots of the launch ejecta velocity v normalized by the impact velocity $v_0$ (upper panel) and of the launch angle given in degree measure unit (lower panel), against the ejecta launch distance from the crater centre normalized by the projectile radius a.}
\label{fig:ve-ang_ejct}
\end{figure}

In order to investigate the dynamical evolution of the ejecta, we analyse the recorded velocity field of the Lagrangian tracers. In Fig.~\ref{fig:ve-ang_ejct}, a summary of the ejecta launch velocity and launch angle against the launch position is provided for the DRPR rheological model. Similar results were obtained also with the LUND model. Irrespective of the model used, boulders are launched at distances from the impact point up to about three times the projectile radius. 
The fastest ejecta are launched close to the impact point, and reach values as high as few times less the initial impact speed. At a projectile diameter from the impact point, the ejecta speed undergoes a kink, with a value of 2-3 tenth the initial impact velocity (cf. Fig.~\ref{fig:ve-ang_ejct}).

\begin{figure}[h!]
\centering
\includegraphics[width=\hsize]{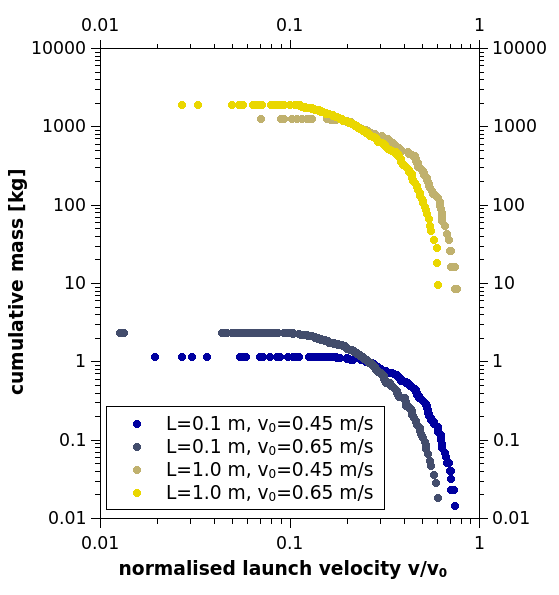}
\caption{Plot of the cumulative (absolute) ejecta mass over normalised velocity. The two sets of curves in the yellow and blue colour tones refer to the two different projectile sizes of 1.0 m and 0.1 m, respectively.}
\label{fig:mass}
\end{figure}

Launch angle values of the boulders are clustered between values of $30^\circ$ and $60^\circ$, with respect to a plane tangential to the impact point, at a launch position between two and three projectile radii from the impact point. Closer to the impact point (less than 2 radii) the angle is between $30^\circ$ and $50^\circ$ and increases with the distance. In contrast, the ejection angles with respect to a plane 
tangential to the impact point of DART were found lower (about $30^\circ$) from the measurements taken by the LUKE camera \citep{Desh}.\\
Assuming each tracer having the mass of the initial toroidal ring it is associated to, we derive a cumulative plot of the total ejected mass against the launch velocity. Once normalized to the initial impact speed, the curves for each of the two tested projectile sizes are similar in shape and intersect at 0.2 the normalized velocity (cf. Fig.~\ref{fig:mass}). It is worth noticing that a factor of three in the ejected mass is found for the two projectile sizes (yellow-like and blue-like colour in Fig.~\ref{fig:mass}).
\\

\section{Results: dynamics of ejecta from sesquinary impacts}

Once the ejecta of a sesquinary impact leave the asteroid surface, they are subject to the gravity of the asteroid pair, the solar radiation pressure, and the solar tide. The physical properties of the ejecta (mass and ejection velocity), derived from the iSALE modelling, are used as initial conditions for a set of numerical integrations with the LICEI code. These are carried out until they escape the double asteroid system or, in the majority of the cases, re--impact on one of the two bodies.

To generate the particles for the dynamical simulation we take the results presented in Fig.~\ref{fig:ve-ang_ejct} and use them to create a representative sample of ejecta particles for different sesquinary impacts. The parameters of the simulated impacts are taken from simulations performed in \cite{langner2024} and are listed in Tab.~\ref{tab::sim1}. In the selection of the specific impacts, 
we focused on events whose velocities are close to those used in iSALE simulations (45 cm/s and 65 cm/s). The exception is the slowest impact number 4 ($\approx 30$cm/s), where the difference is greater, but we decided to use it as it can represent slower impacts. The iSALE impact simulation assumes cylindrical symmetry; therefore, for each point in Fig. \ref{fig:ve-ang_ejct} we generate a number of fictitious particles, proportional to the mass of the points from the iSALE simulation. The initial conditions are then randomly rotated around the assumed impact point. The ejection angle and initial position with respect to the impact point are taken directly from the iSALE results. In order to generate the initial velocities, we use the relative values of velocity shown in Fig.~\ref{fig:ve-ang_ejct} and multiply them by the velocity of the specific impact from Tab.\ref{tab::sim1}. The total number of particles we generate for each impact is about 2500, and for each of them we assign a random size between 1 mm and 1 cm using a power law with exponent $-2.8$. The upper limit of 1 cm comes from the ejected mass estimation from iSALE results. If particles grater than 1 cm are ejected they should be rare and do not have any effect on the overall results of the simulation. For the lower bound we decided to not include sub-millimeter sized particles. We note that the Solar radiation pressure would have a larger effect on these very small objects, making their orbits less stable (they will re-impact or escape earlier), with no overall effects on the results.
According to our N--body simulations, most of the ejected particles re--impact on Didymos without completing a single orbit around the asteroid. A smaller fraction of particles is injected in almost stable orbits and remains in the system (up to about 40 days). In the end, such particles either collide with Didymos or Dimorphos or escape on hyperbolic trajectories from the system.

\begin{table}[h!]
\caption{Characteristics of the sesquinary impacts on Didymos used in the simulations: epoch (in days after the DART impact), velocity of the impact, location of the event on the surface of the body (latitude and longitude).}
\label{tab::sim1}
\centering
\begin{tabular}{l|c c c c }
\hline\hline
& time & velocity & latitude & longitude \\
& (days) & (cm/s) & (deg) &(deg) \\
 \hline
1. & 108.250000	& 40.5443 & 23.529223 &	-15.279906\\
2. & 150.416667	& 72.2430 & -10.775361 & 51.902415\\
3. & 178.583334	& 67.5132 & -25.374971 & 4.331503\\
4. & 197.000000	& 30.4752 & 10.938186 &	146.574654\\
5. & 232.333334	& 72.7596 & -10.354222 & 93.068402\\
\hline
\end{tabular}
\end{table}

\subsection{Ejecta lifetime vs. asteroid orbit}

It is noteworthy that the fraction of ejecta remaining in the system varies among different sesquinary impacts. The behaviour is strongly dependent on the velocity and geometry of the impact and on the position and distance of the Sun with respect to the location of the impact. The solar radiation pressure increases the orbital eccentricity of the ejected dust, lowering in this way the pericentre distance resulting in a collision with the surface of Didymos. 

\begin{figure}
   \centering
   \includegraphics[width=0.45\hsize]{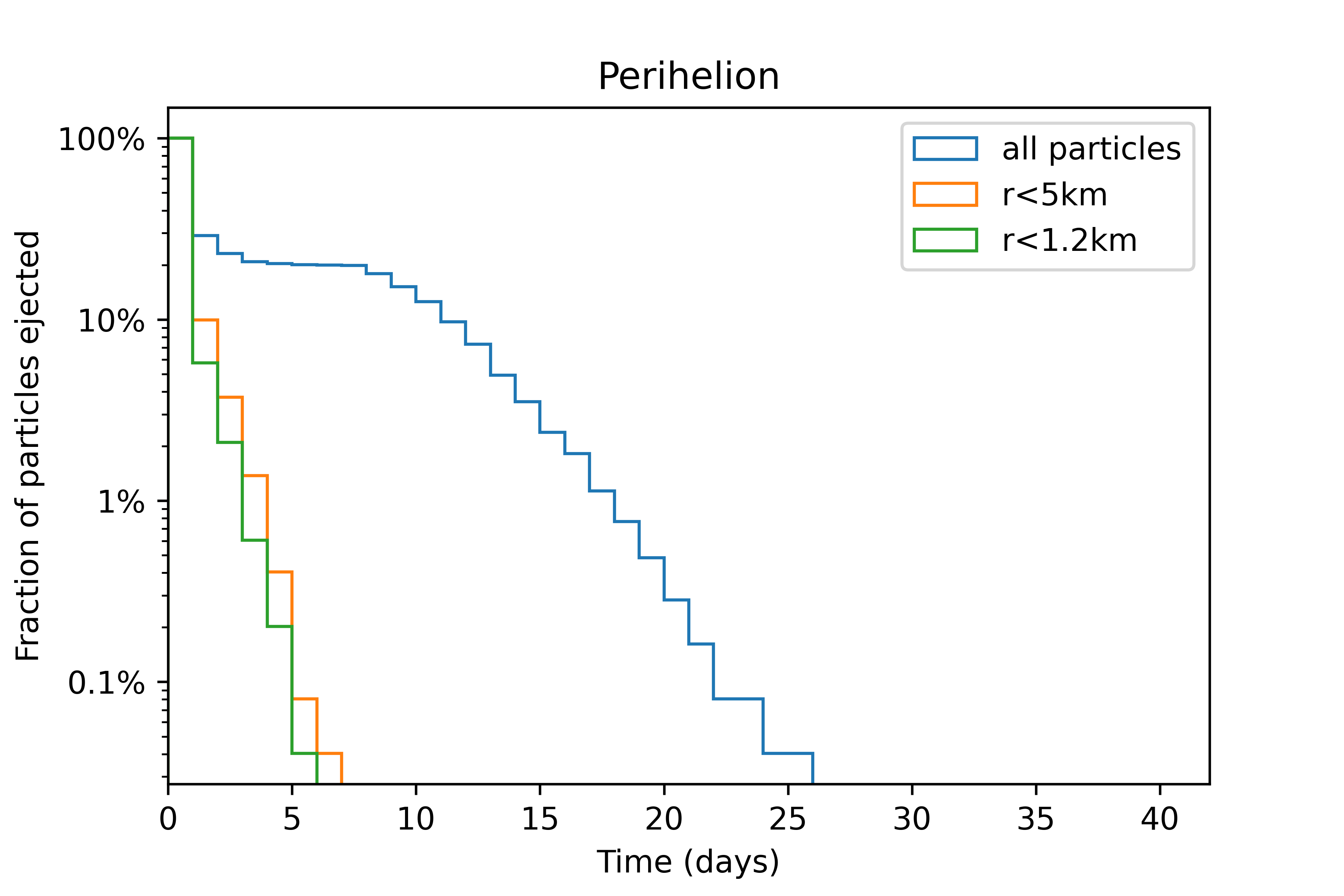}
   \includegraphics[width=0.45\hsize]{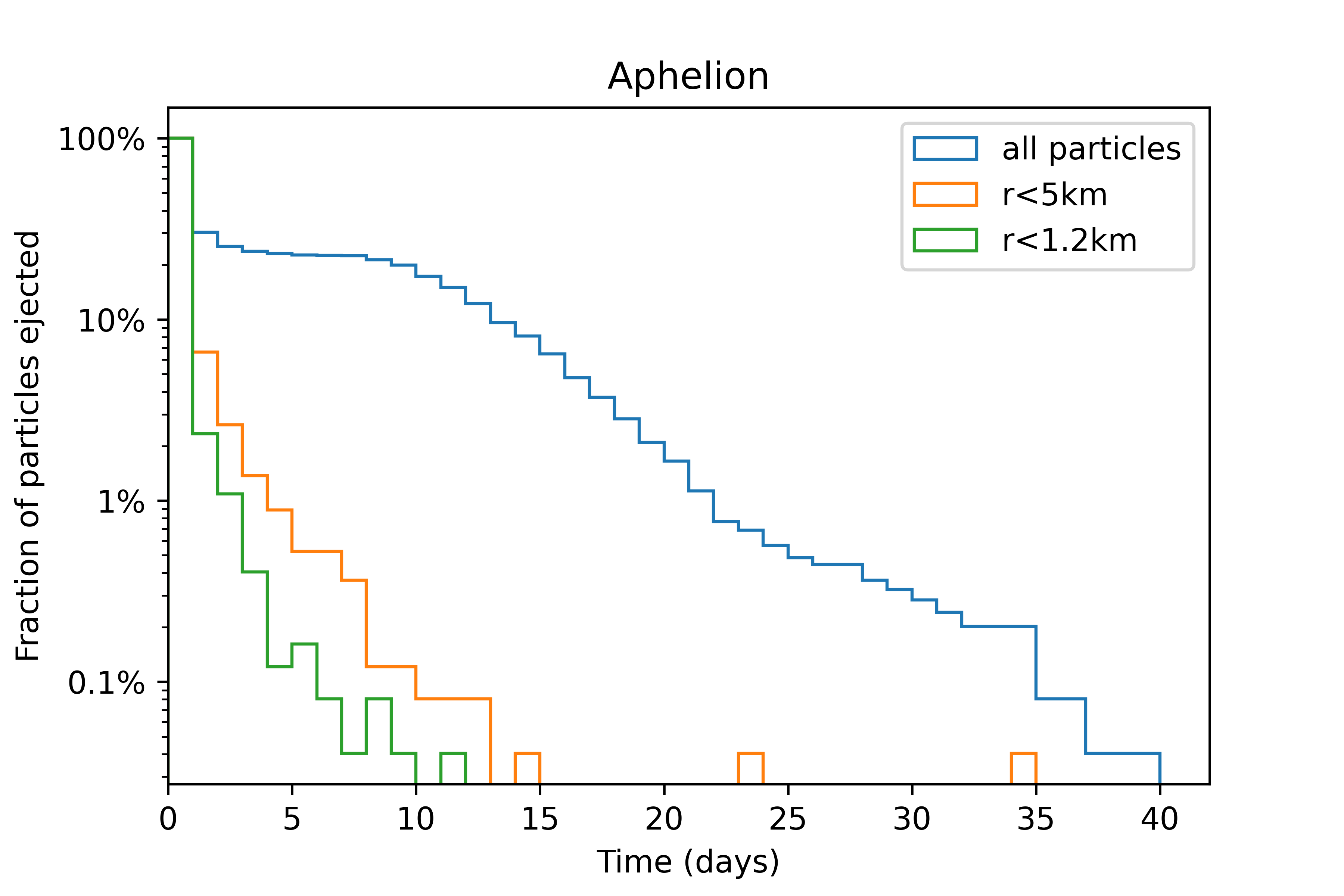}
      \caption{The number of particles in the system after late impact when Didymos is near its perihelion (left) and aphelion(right). Colour bars represents the number of particles in specific range of a distance from Didymos barycentre ($r$). Almost all particles are removed after 15 (perihelion case) or 30 days (aphelion), and the close proximity of Didymos is free of particles after 5--10 days. }
      \label{fig::peri} 
\end{figure}

The dependence on the solar radiation pressure is clearly visible when comparing the results of two simulations for the same sesquinary impact (thus using the same iSALE-derived parameters for the ejecta), but one assumed to occur near the Didymos--Dimorphos perihelion and the other one when the asteroid pair is near the aphelion. 
In the case of an impact at the perihelion, the typical lifetime of the ejecta is short, and after 5 days the area in the closest proximity to Didymos is almost cleared of dust particles (fig. \ref{fig::peri}). In the aphelion case, the lifetime is longer and most of the particles escape the system or re--impact on the surface within approximately 15 days after the impact. The shorter lifetime is also the effect of stronger solar gravity perturbations, but they become significant when the particle is already at larger distance from Didymos.

\subsection{Contribution of sesquinary impacts to a dust tail}
\label{sec:tail}
\begin{figure}
   \centering
   \includegraphics[width=\hsize]{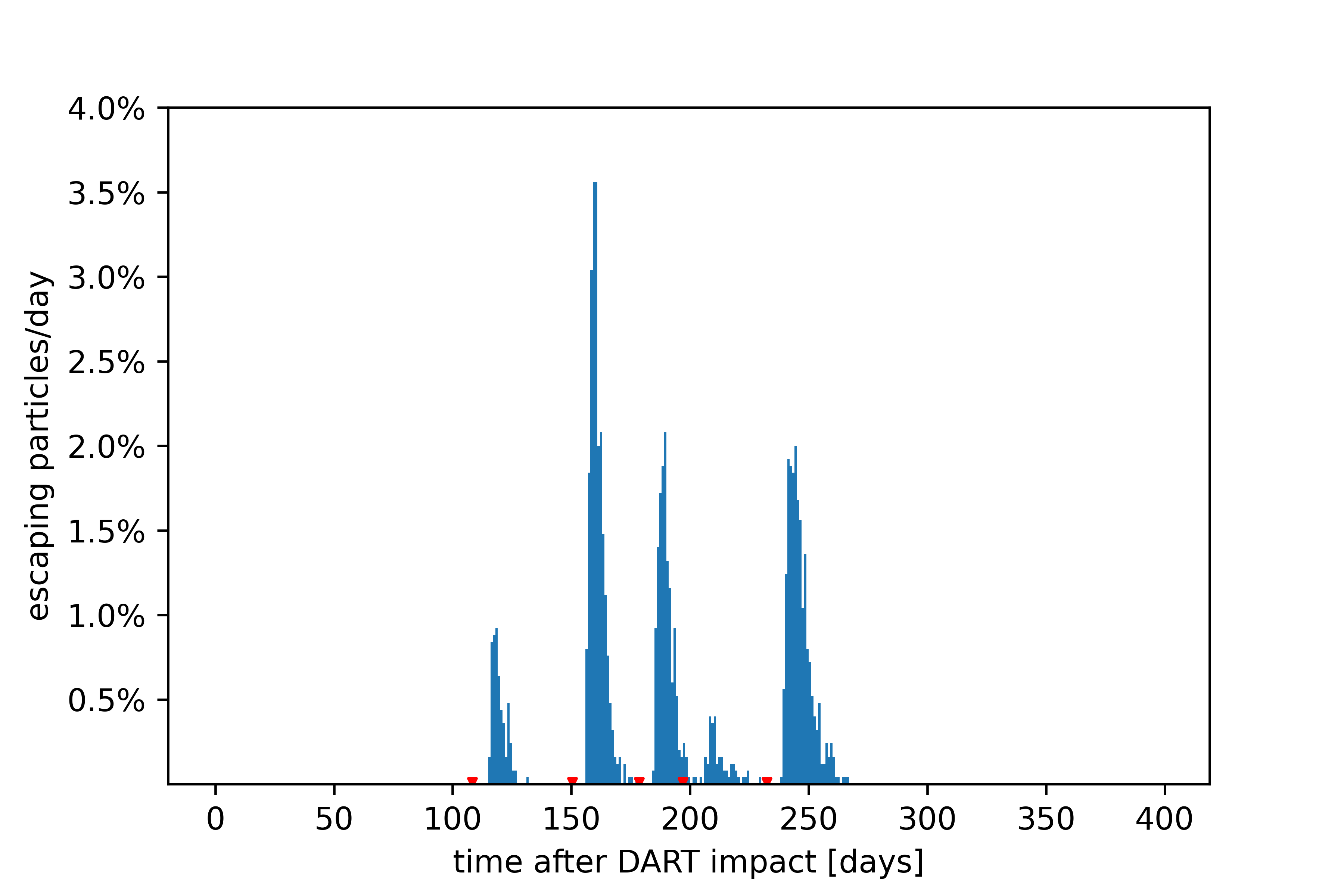}
      \caption{The escaping particles from 5 sesquinary impacts. For each impacts we generated about 2500 sample particles. The x-axis represents the time measured from the DART impact and red triangles corresponds to the simulated sesquinary impacts time. The total percentage of escaping objects in the 5 histograms are: 5.4\%, 22.0\%, 13.6\%, 2.7\% and 19.8\%. In all the cases displayed here, the escaping particles reinforces the tail for up to 20-30 days after the impact.}
\label{fig::escapes}
\end{figure}

\begin{figure}
   \centering   \includegraphics[width=\hsize]{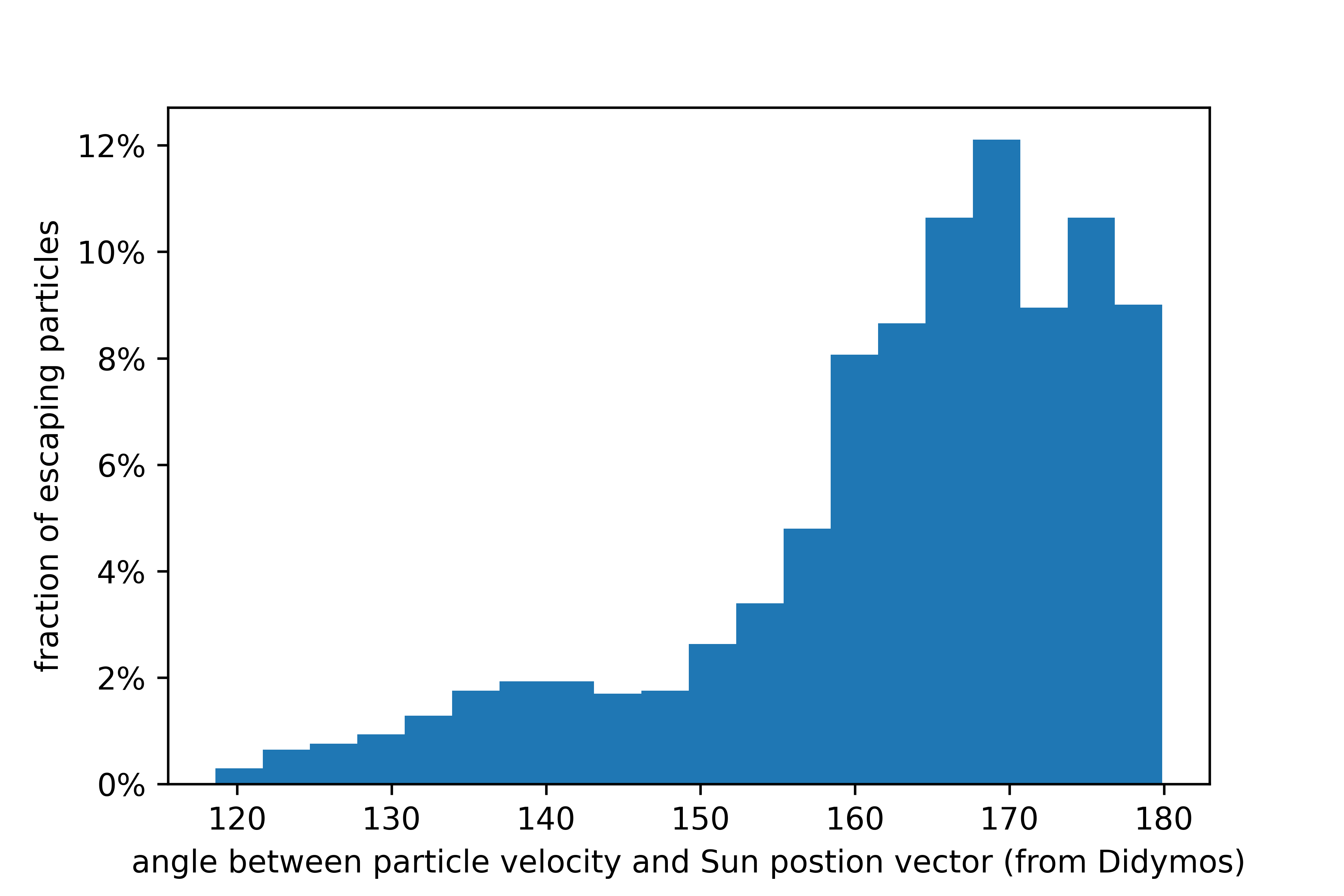}
      \caption{Distribution of the angle between the escaping particles velocity vector and the Sun direction. As mentioned in the text, all the particles tend to escape in the direction opposite to the Sun.}
         \label{fig::escape_dir} 
\end{figure}

The dynamical simulations show that the number of escaping particles (Fig.~\ref{fig::escapes}) is not sufficient to provide a continuous flow of dust that may lead to an extended dust tail. Each boulder impact creates a spike that, by itself, would be a short living feature, but it might contribute to sustain a possible pre--existing tail. Indeed, as shown in Fig.~\ref{fig::escapes}, a clear correlation between the percentage of escaping particles and the velocity of the impact is apparent. The largest spikes, in terms of percentage of escaping particles, are related to the impacts number 2, 3, and 5 in Tab.~\ref{tab::sim1}, with the low velocity impact number 4 ($\sim 30$ cm/s) being clearly the smallest one. Irrespective of the impact velocity, in all the histograms a similar behaviour is shown, with the majority of particles escaping in the aftermath of the impact, followed by a steep decrease with a full clearing within a few weeks after the event. All particles after the impact escape in the direction opposite to the Sun (as shown in Fig. \ref{fig::escape_dir}), confirming the key role of the solar radiation pressure in the escape mechanism.

\subsection{Maps of re--impact sites}

\begin{figure}
   \centering
   \includegraphics[width=\hsize]{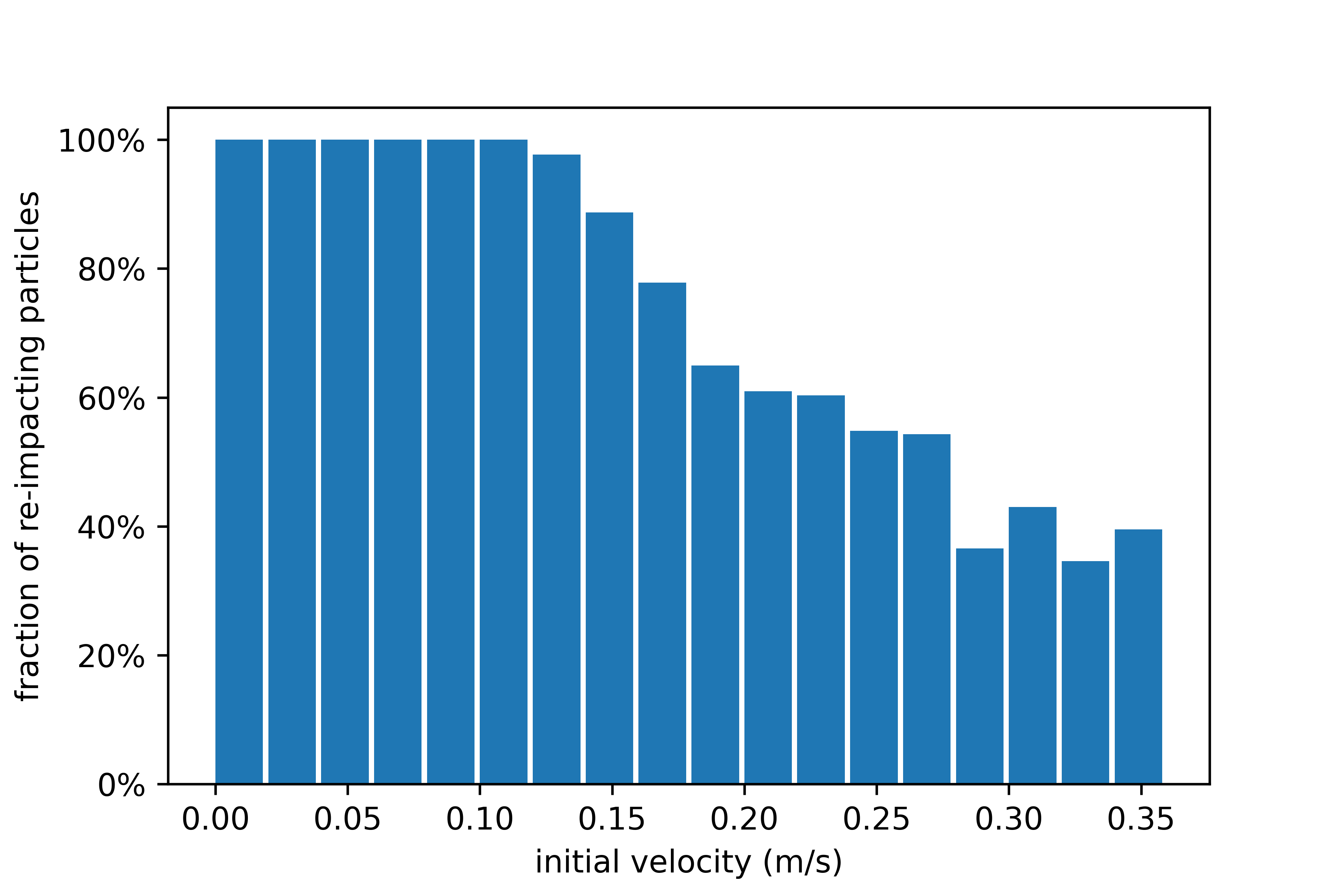}
      \caption{Percentage of re-impacting particles, as a function of their initial velocities, for all the impacts listed in Tab.~\ref{tab::sim1}. Particles that are not re-impacting escape the system. The minimum particle velocity required to  escape from the system is around 0.1 m/s.}
\label{fig::reimpact_percents}
\end{figure}

As described in Sec.~\ref{sec:tail}, only a small fraction of the ejecta escapes and possibly contributes to a dust tail according to our modelling. The majority of the dust raised by sesquinary collisions re--impacts on Didymos. In Fig.\ref{fig::reimpact_percents}, we show the percentage of re--impacting particles depending on their initial velocity. This shows that all the slower particles as well as more than 50\% of the faster ones re-impacts Didymos. 
The lower bound for impact velocity to produce significant population of non re--impacting particles is between 30 and 40 cm/s.  

\begin{figure*}
   \centering \includegraphics[width=0.32\hsize]{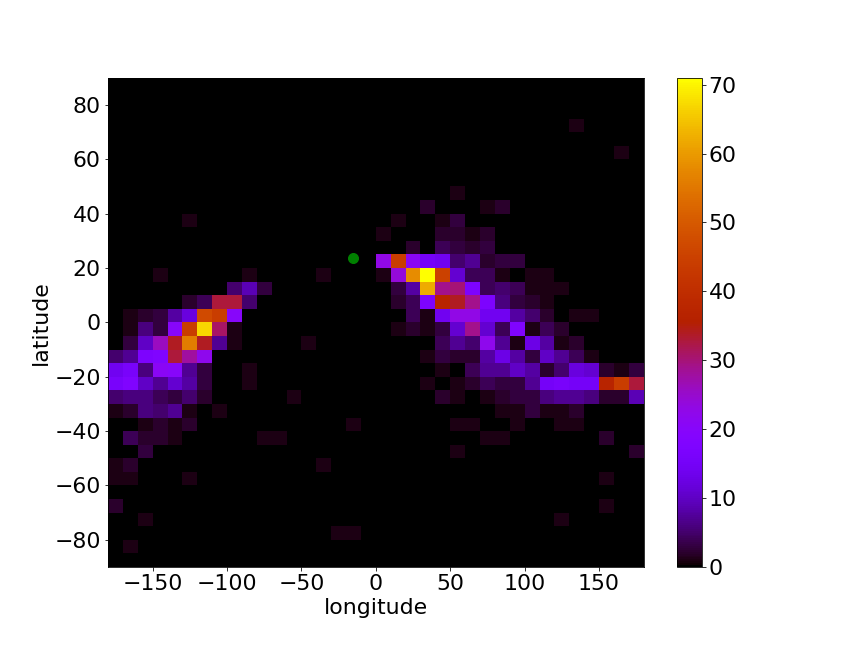}
   \includegraphics[width=0.32\hsize]{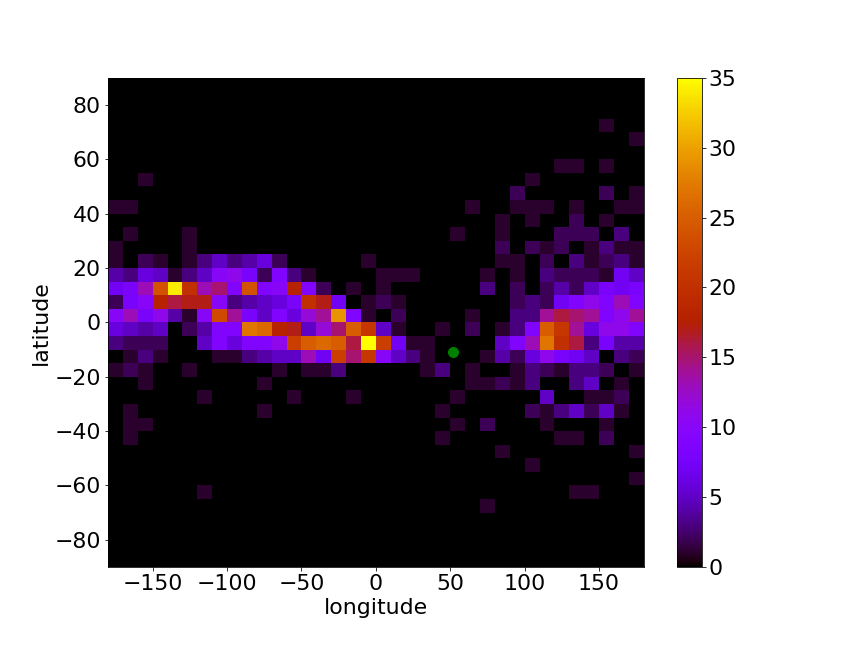}
   \includegraphics[width=0.32\hsize]{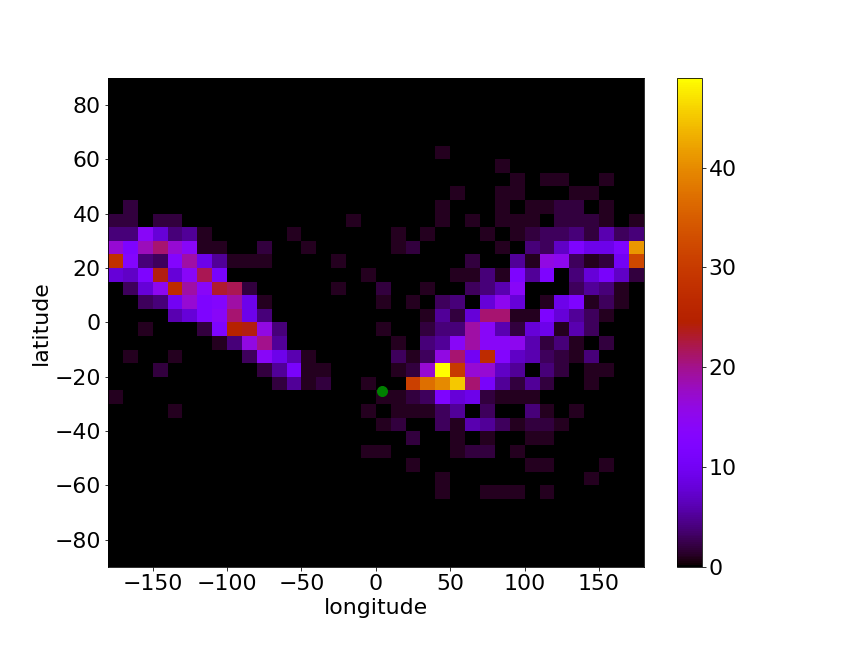}\\
   \includegraphics[width=0.32\hsize]{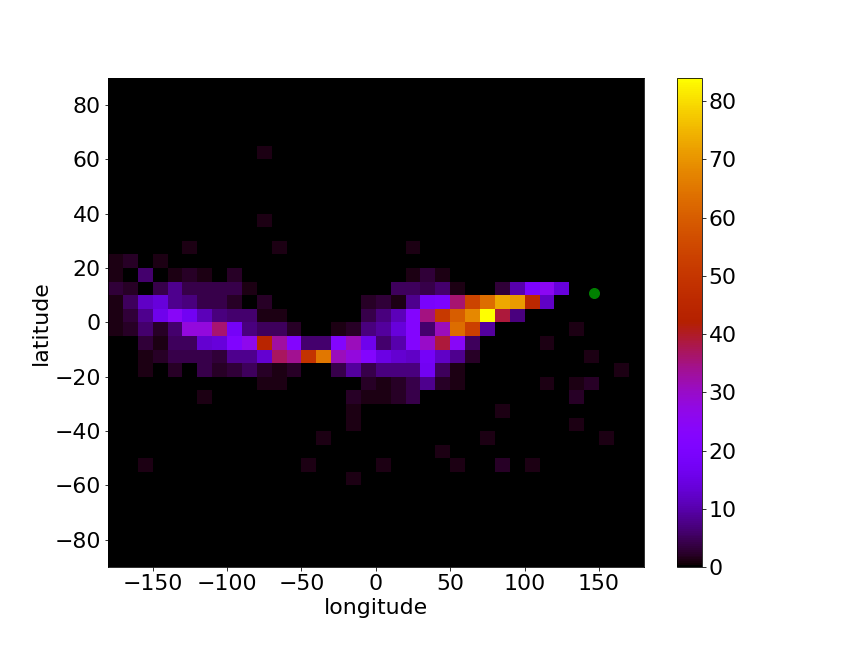}
   \includegraphics[width=0.32\hsize]{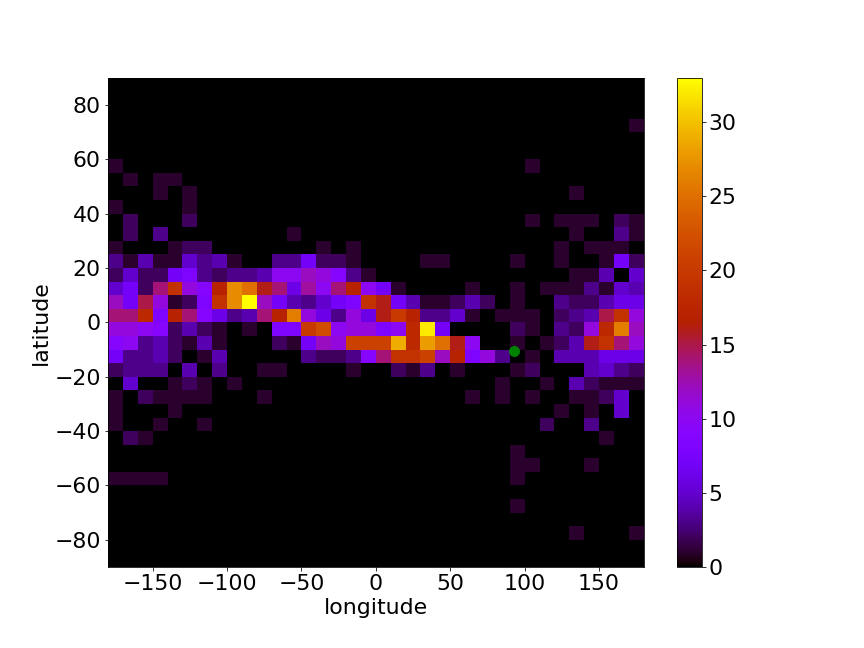}
   \includegraphics[width=0.32\hsize]{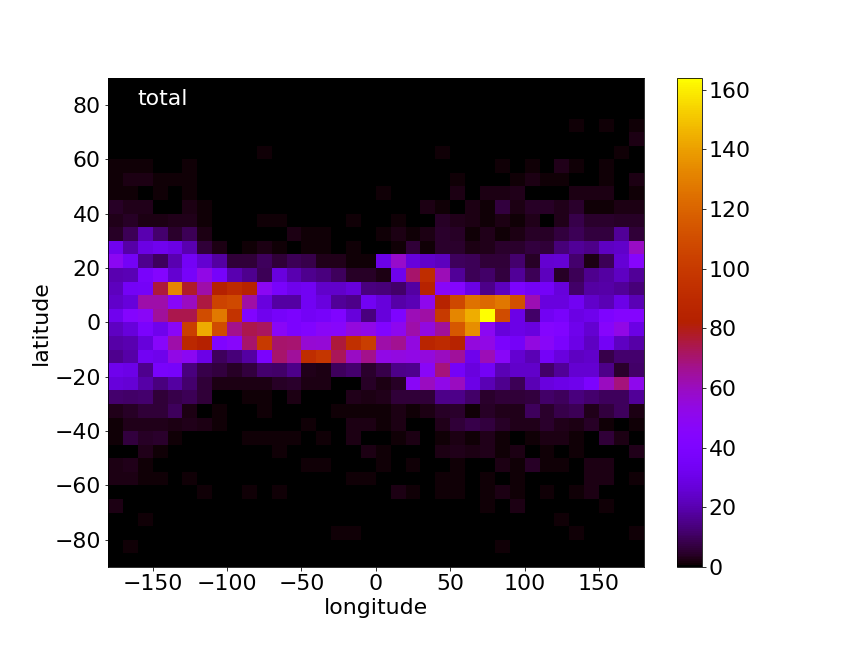}
      \caption{Density of particles colliding against Didymos produced by the sesquinary impacts described in Tab.~\ref{tab::sim1}. The single green dot represents the location of the boulder impact. In each case the dust is scattered over large areas on the surface. The shape of this area can vary between each event. In the bottom right panel we combine all the 5 previous  distributions of dust particles for the boulder impacts into a single 2D histogram.}
      \label{fig::reimpacts_density} 
\end{figure*}

From the outcome of our simulations we can draw maps of the re--impact locations on the surface of Didymos in the form of 2D histograms (Fig.\ref{fig::reimpacts_density}).
The dust is usually scattered on large areas on the surface of the asteroid and the pattern depends on the individual impact. It is difficult to find a simple way to correlate the initial impact conditions with the final distribution of re--impacting fragments. Due to the critical rotation velocity of Didymos, even dust particles with low ejection velocity would achieve sub--orbital trajectories and travel a significant distance before landing back on the surface, especially if they are ejected in the asteroid spin direction. The bottom right 2D histogram in Fig.\ref{fig::reimpacts_density} shows the total distribution of re--impacting ejecta from five simulated sesquinary impacts, showing that the dust covers most of the near--equator area of Didymos. Due to the mostly equatorial location of the re--impacts on Didymos, the dust patterns are usually confined to latitudes below $\pm 40$ deg. 
The map shown in Fig.\ref{fig::reimpacts_density} will therefore be instrumental in the interpretation of the surface properties as detected by Hera, in case any significant difference would be found between lower and higher latitudes terrains.

A small number of particles raised by the sesquinary impacts are able to leave Didymos and impact on Dimorphos. However, this tiny flux is not able to account for a significant mass transfer between the two asteroids. In most of the simulations it is less than 0.1\% of the simulated particles. Therefore the mass transfer between the two asteroids due to sesquinary impacts is not significant. However, Dimorphos can have a significant impact on the orbits of ejected dust. In some cases the fastest particles from the ejecta cone created by sesquinary impacts on Didymos can fly relatively close to the surface of Dimorphos. Then the gravitational pull from Dimorphos alters the trajectory of those ejecta and can increase their pericentre or apocentre distance. This effect can increase the number of particles that survive their first orbital period and, therefore, can increase the fraction of particles that escape the system. 

\section{Discussion and conclusions}

Numerical simulations of the dynamics of ejecta from the DART impact show that m--sized boulders can impact the surface of both Didymos and Dimorphos at later times, extending up to 400 days after the DART event. These impacts, termed sesquinary, have very low velocity, but they may have important consequences for the evolution of the system. We have performed simulations using the iSALE shock physics code to find the mass and ejection velocity of the ejecta of representative sesquinary impacts on Didymos. We find that most of the particles between 1 mm and 1 cm in size can depart from the asteroids' surface with velocities larger than 10\% the impact velocities (Fig.~\ref{fig:ve-ang_ejct} top). The LICEI code is fed with these values and a large number of potential ejecta from the sesquinary impacts (mainly against Didymos) is numerically integrated in time. We find that the vast majority of the ejecta re--impact on the surface of the asteroid after approximately 5 to 10 days (depending on the asteroid location along its heliocentric orbit) after the impact. During this period they create a dusty haze around the asteroid and then re--impact in different locations of the surface of the asteroid contaminating the local surface composition. 

A smaller fraction of the ejecta escape contributing to a dusty tail. However, their contribution is in the form of short living spikes that may help to sustain a previously existing dust tail but cannot account for an independent one. In particular, some of these secondary impacts, which should have been particularly frequent in the initial weeks after the DART impact, could have been responsible for the puzzling long term outliving of the observed tail \citep{NatureLi}.  

Due to short lifetime of the dust and the rather small probability of larger boulders to survive in the Didymos-Dimorphos system until the Hera arrival \citep{langner2024}, it is very unlikely that this dust will be present on the orbit and should not be considered as a threat to Hera mission. However, the impacts should have altered the surface of Didymos and should be considered during the analysis of the data received from the mission. Considering the resolution of the Asteroid Framing Cameras instrument on Hera \citep{Michel}, the small and shallow craters produced by sesquinary impacts will not be observable until the close operation phase of the mission when the resolution becomes better than 1 m/pixel. Even then their size will be at most a few pixels in diameter. It is worth stressing that, even though we simulated impacts using 1 m projectiles (that should be more frequent in the system), the boulders observed by \cite{Jewitt} include also larger objects of several meters in size. 
A deep analysis of the scaling of the craters diameter between different impactors sizes is beyond the scope of this paper, given also the low number of simulations performed. However, from the available data, we note that the craters produced by 1 m objects should scale almost linearly to larger projectiles such as those observed by \cite{Jewitt}, hence producing craters of several meters in diameters, that could be more easily detectable.

The fresh dust released by the impact is going to settle on the surface of Didymos mostly closer to the equator. This fresher material, excavated down to a meter below the surface can result in a difference in the observed surface spectral properties of the asteroid between the lower and higher latitudes.

\begin{acknowledgements} 
This research was supported by the Italian Space Agency (ASI) within the  Hera Projects (2022-8-HH.0, “Missione HERA – Attività scientifiche per la Missione HERA”). 
R. L. acknowledges the funding from ESA, project S1-PD-08.2.
We gratefully acknowledge the developers of iSALE‐2D/Dellen version (https://isale-code.github.io), including Gareth Collins, Kai Wünnemann, Dirk Elbeshausen, Tom Davison, Boris Ivanov, and Jay Melosh. Some plots in this work were created with the pySALEPlot tool written by Tom Davison.
\end{acknowledgements}
\bibliographystyle{aa} 
\bibliography{biblio.bib} 
\begin{appendix}
\section{iSALE model parameters}
\begin{table}[h!]
\caption{iSALE model parameters}
\label{tab::isale-input}
\centering
\begin{tabular}{p{1.5cm} p{3.5cm} p{1.5cm}}

\hline\hline
Symbol & Description  & Value \\
\hline
$Y_0$ & Damage strength at zero pressure	 & 10.0 Pa\\
\\
$f$ & Internal friction coefficient (damaged)	 & 0.6\\
\\
$Y_M$ & Strength at infinite pressure	 & 1.0 GPa\\
\\
$\phi$. & porosity	 & 40\%, 60\% \\
\\
\hline
\\
Eos &   & Tillotson\\
\\
Strength model & 	 & DRPR, LUND\\
\hline
\end{tabular}
\end{table}
\end{appendix}
\end{document}